\documentclass[a4paper,11pt]{article}
\pdfoutput=1 

\usepackage{jheppub} 

\usepackage[T1]{fontenc} 
\usepackage{bm}
\usepackage{listings}
\usepackage{xcolor}
\usepackage[export]{adjustbox}
\usepackage{caption}
\usepackage{subcaption}

\usepackage{fast_likelihoods-defs}

\title{\boldmath Simplified likelihoods using linearized systematic uncertainties}

\author[a]{N. Berger\note{Corresponding author.}}

\affiliation[a]{LAPP, Univ. Savoie Mont Blanc, CNRS/IN2P3, Annecy}

\emailAdd{nicolas.berger@cern.ch}

\abstract{This paper presents a simplified likelihood framework designed to facilitate the reuse, reinterpretation and combination of LHC experimental results. The framework is based on the same underlying structure as the widely used \HistFactory\ format, but with systematic uncertainties considered at linear order only. This simplification leads to large gains in computing performance for the evaluation and maximization of the likelihood function, compared to the original statistical model. The framework accurately describes non-Gaussian effects from low event counts, as well as correlated uncertainties in combinations. While primarily targeted towards binned descriptions of the data, it is also applicable to unbinned models.}

\begin{document} 

\maketitle
\flushbottom

\section{Introduction}
\label{sec:intro}

The statistical models describing experimental measurements are a key component of LHC data analysis. Consisting of the probability distribution function (PDF) of the measurement together with the observed dataset, they are used to compute the final experimental results --- e.g. confidence intervals for model parameters, or significance values for possible excesses over background --- often through the use of frequentist profile-likelihood ratio (PLR) methods~\cite{Asimov}. They can also be utilized to make further use of the measurement information, for instance in combinations with other results, or as reinterpretations in the context of alternative signal models.

Despite this central role, statistical models are not systematically made available as part of experimental publications. This is partly for technical reasons: first, they are often complex, with up to $O(10^4)$ parameters in some cases~\cite{CMS:2022dwd}. A single maximization the likelihood function, which is needed to compute the PLR, can therefore require up to several hours or days of computation time. Another limitation is the fact that the statistical models of LHC measurements are typically implemented within formats and tools not widely used in other fields, such as the \ROOT\ framework~\cite{ROOT}.

The information provided in publications, such as the best-fit value of the parameters of interest (POIs) and the covariance matrix of their measurement, typically allow a partial reconstruction of the model. However this is only possible under additional assumptions -- in particular Gaussian approximations that do not accurately describe data taken in the Poisson regime with low expected event counts. In cases where full PLR scans are published, the description of systematic uncertainties also does not typically allow a full separation of the different sources of uncertainty, so that correlations across different measurements cannot be properly accounted for when performing their combination.

For these reasons, recent efforts have encouraged the publication of faithful representations of the experimental statistical models under FAIR (Findable, Accessible, Interoperable, Reusable) principles~\cite{FAIR}, in particular with a view towards reinterpretations targeting alternative physics models~\cite{RIF1,RIF2}. This objective can be realized in particular through their publication in open formats. Some recent progress has been achieved in this direction, such as the publication of statistical models by the ATLAS collaboration using the \pyhf~\cite{pyhf_joss,pyhf_soft} framework.
These cases however remain rare so far, in particular due to the limitations described above.

Simplified likelihoods offer compromise solutions that aim to provide less complex descriptions of the experimental statistical models that remain more accurate than Gaussian models. Several approaches have been proposed~\cite{Buckley:2018vdr, Collaboration:2242860, Coccaro:2019lgs, Fichet:2016gvx, Cranmer:2013hia}. This work describes a simplification applied to statistical models, in which  
the dependence on the POIs of the measurement is treated exactly, but the remaining \emph{nuisance parameters} (NPs) are considered at linear order only. This allows the maximization of the likelihood function with respect to the NPs (usually denoted as \emph{profiling} the likelihood function) to be performed in closed form using matrix algebra techniques. This in turn can significantly decrease the computing times of the PLR computation since the NPs, which are used to describe in particular systematic uncertainties, typically form a large fraction of the model parameters. The structure of the simplified model, in terms of the POIs, NPs, measurement regions and event samples, remains faithful to the original model. The models are stored in plain text, and computations are performed using python-based tools.
The method is applicable to both binned and unbinned descriptions of the experimental data, with unbinned models treated in a binned approximation. This flavor of simplified likelihoods is denoted as Simplified Likelihoods with Linearized Systematics (SLLS) in the rest of this paper to avoid confusion with other simplified likelihood formats.

The paper is organized as follows: the SLLS formalism is presented in detail in Section~\ref{sec:formalism}; Section~\ref{sec:trileptons} shows a realistic application to a ATLAS search for supersymmetric particles; an application to an unbinned model is presented in Section~\ref{sec:unbinned}, and Sections~\ref{sec:discussion} and~\ref{sec:conclusion} present a discussion of these results and conclusions.

\section{Simplified likelihood formalism}
\label{sec:formalism}

\subsection{The \HistFactory\ framework}
\label{sec:histfactory}

The simplified likelihoods described in this work are based on the \HistFactory\ framework~\cite{HistFactory}, which is widely used in LHC experiments and implemented within both \ROOT\ and \pyhf.

It encodes measurements derived from multiple event counts as a set of \emph{channels}, each corresponding to an independent set of data, consisting of one or several counting bins.
In each bin, a counting experiment is described using a Poisson distribution. Each expected event yield is expressed as a sum of contributions from several \emph{samples}, representing both signal(s) and background(s), and each is a function of the POIs and NPs of the model. Systematic uncertainties are represented as NPs that are constrained by external information described by a constraint PDF. This constraint is a representation of a separate \emph{auxiliary} experiment, sensitive to the value of the NP through the measurement of an \emph{auxiliary observable}. The full likelihood function is written as
\begin{equation}
    L(\vm, \vt) =
    \prod\limits_{c=1}^{N_{\text{channels}}}
    \prod\limits_{b=1}^{N_{\text{bins}, c}}
    \text{Pois}\left(n_{cb}, \sum\limits_{s=1}^{N_{\text{samples},c}} \nu_{cbs}(\vm,\vt) \right) 
    \prod\limits_{l=1}^{N_{\text{constraints}}} C_l(\tilt_l, \theta_l)
    \label{eq:L}
\end{equation}
where the index $c$ runs over the $N_{\text{channels}}$ measurement channels, $b$ runs over the $N_{\text{bins}, c}$ bins in channel $c$, and $s$ over the $N_{\text{samples},c}$ samples. The  observed event yield in bin $b$ of channel $c$, denoted by $n_{cb}$, is described by the Poisson PDF Pois in terms of the expected yields $\nu_{cbs}(\vm,\vt)$ for each sample $s$. The \vm\ and \vt\ refer collectively to the POIs and the NPs, respectively, and the index $l$ runs over the $N_{\text{constraints}}$ constrained NPs $\theta_l$ and their respective auxiliary observables $\tilt_l$.
The constraints $C_l$ are in principle arbitrary but in practice either Poisson or Gaussian forms are used, depending on the properties of the associated systematic uncertainty. 

\subsection{Simplified likelihoods with linearized systematics}
\label{sec:linearization}

The SLLS formalism introduced in this paper brings two simplifications to the \HistFactory\ description. Firstly, the impact of the NPs on the log-likelihood value is described at linear order only. In particular, the $\nu_{cbs}$ are expressed as a linear function of the NPs,
\begin{equation}
    \nu_{cbs}(\vm, \vt) = \nu_{cbs}\nom(\vm) \left[ 1 + \sum\limits_{k=1}^{N_{\text{NP}}}\Delta_{cbsk}(\theta_k - \theta\nom_k) \right].
    \label{eq:nexp_linear}
\end{equation}
The $\nu_{cbs}\nom(\vm)$ are the expected event yields computed at the nominal values $\theta\nom_k$ of the NPs. The $\Delta_{cbsk}$ are linear coefficients specifying the impact of $\theta_k$ on $\nu_{cbs}$, for each of the $N_{\text{NP}}$ parameters $\theta_k$. As noted above, the dependence of the $\nu_{cbs}(\vm, \vt)$ on the parameters of interest \vm\ is described exactly.
The linear approximation in the impact of the NPs is also applied to the Poisson distributions, as described in Appendix~\ref{app:linearization}.
Secondly, the constraints $C_l$ are all assumed to be Gaussian, and are collectively represented as a single multivariate Gaussian PDF with central value $\tilde{\vt}$ and inverse covariance matrix $\Gamma$.

With these assumptions, the profiled value $\hat{\hat{\theta}}_k(\vm) = \argmax_{\theta_k} L(\vm, \vt)$ of the parameter $\theta_k$ at a given value \vm\ of the POIs can be computed in closed form as

\begin{equation}
\hat{\hat{\theta}}_k(\vm) = \theta\nom_k + \sum\limits_{k'}\left[(\Gamma + P(\vm))^{-1}\right]_{kk'} \left[\sum\limits_{k''} \Gamma_{k'k''}(\tilde{\theta}_{k''} - \theta\nom_{k''}) - Q_{k'}(\vm)\right].
    \label{eq:profile}
\end{equation}
with the vector $Q(\mu)$ and the matrix $P(\mu)$ given by
\begin{align}
Q_k(\vm) &= \sum\limits_{c=1}^{N_{\text{channels}}} \sum\limits_{b=1}^{N_{\text{bins}, c}}
\left(\nu_{cb}\nom(\vm) -  n_{cb} \right) \sum\limits_{s=1}^{N_{\text{samples},c}}\frac{\nu_{cbs}\nom(\vm)}{\nu_{cb}\nom(\vm)}   \Delta_{cbsk} \\
P_{kk'}(\vm) &= \sum\limits_{c=1}^{N_{\text{channels}}} \sum\limits_{b=1}^{N_{\text{bins}, c}}
n_{cb} \sum\limits_{s,s'=1}^{N_{\text{samples},c}}  \frac{\nu_{cbs}\nom(\vm) \nu_{cbs'}\nom(\vm)}{[\nu_{cb}\nom(\vm)]^2} \Delta_{cbsk} \Delta_{cbs'k'}
    \label{eq:pq}
\end{align}
where $\nu_{cb}\nom(\vm) = \sum_s \nu_{cbs}\nom(\vm)$.
The $Q_k$ and $P_{kk'}$ terms in Eq.~\ref{eq:profile} encode the impact of the data on $\hat{\hat{\theta}}_k(\vm)$, while the terms involving $\Gamma_{kk'}$ originate from the constraint PDFs in the likelihood function. While $P_{kk'}$ is quadratic in the $\Delta_{cbsk}$, it generally cannot be neglected, in particular in the case of NPs that are not associated with a constraint PDF.

Using these relations, the profiling of the NPs at a given \vm\ can be performed using simple matrix algebra. The sizes of the matrices is given by the number of NPs, which can be fairly large -- in some cases up to $O(10^4)$ -- but building these matrices and performing multiplication and inversion operations is nevertheless far quicker than the non-linear maximization of the full likelihood function using e.g. a gradient descent algorithm.

While the form given in Eq.~\ref{eq:nexp_linear} is used to profile the NPs, the evaluation of the likelihood function uses instead the alternative form 
\begin{equation}
\nu_{cbs}(\vm, \vt) = \nu_{cbs}\nom(\vm) \exp\left[ \sum\limits_{k=1}^{N_{\text{NP}}}\Delta_{cbsk}(\theta_k - \theta\nom_k) \right].
\label{eq:nexp_non_linear}
\end{equation}
This guarantees that $\nu_{cbs}(\vm, \vt) \ge 0$ for all \vt\ as required for the expected event yield of a Poisson PDF, and provides a suitable approximation to Eq.~\ref{eq:nexp_linear} for small values of $\left|\theta_k - \theta\nom_k\right|$ since the two forms are equal at leading order in this quantity.\footnote{Alternatively, a variation of Eq.~\ref{eq:nexp_linear} with a truncation applied to avoid negative $\nu_{cbs}$ can also be used.}

\subsection{Implementation and storage format}
\label{sec:implementation}

A python implementation of the SLLS formalism is provided in the \fastprof\ public package\footnote{\url{https://github.com/fastprof-hep/fastprof}}. It describes the full statistical model, including both the PDF of the measurement and the observed data. It includes tools to evaluate and profile the likelihood function and perform maximum-likelihood fits as well as higher-level computations such for hypothesis testing, limit-setting and confidence interval estimation. Other tools are provided to validate the simplified models and perform other operations such as combining or pruning models. The computations make use of the linear algebra routines included in \numpy~\cite{numpy} and the minimization routines provided by \scipy~\cite{scipy}. 

The statistical models are stored in a plain-text format using the JSON markup language.  The format specifies the POIs, NPs, auxiliary observables, and measurement channels. Each channel is described as a list of samples, specified by the nominal expected bin yields $N\nom_{cbs}$, the linear impacts $\Delta_{cbsk}$ of each NP on the expected yields, and an optional normalization factor $K(\vm)$ that can be an arbitrary function of the \vm. The expected yields are then expressed as in Eq.~\ref{eq:nexp_linear}, with $\nu\nom_{cbs}(\vm) = N\nom_{cbs} K(\vm)/K(\vm\nom)$, where $\vm\nom$ is the value of the POIs for which the nominal yields $N\nom_{cbs}$ are provided.
The format also specifies the observed data, in terms of the observed yields for each bin of each channel and the observed values of the auxiliary observables.

\subsection{Example}
\label{sec:simple_example}

As an illustration, we consider a simple example measurement consisting of a single-bin counting experiment in the presence of both signal and background contributions. The expected background yield is $b_0=1$, with a relative uncertainty $\epsilon=25\%$. The background yield is treated as a NP in the fit, associated with a Gaussian constraint with an auxiliary observable $\tilde{b}$ (as would occur in the case where the background is determined from a control region with a sufficiently large number of events). The observed yield is $n=2$. The JSON specification of the statistical model is given in Figure~\ref{fig:simple_model:listing}.
\begin{figure}[hbtp]
    \begin{lstlisting}[language=json]
{
   "model": {
      "name": "simple_bkg_uncertainty",
      "POIs": [
         { "name" : "xs_signal", "unit" : "fb", "min_value": 0, "max_value": 10, "initial_value": 1 }
      ],
      "NPs": [
         { "name": "np_bkg", "nominal_value": 0, "constraint": 1, "aux_obs": "aux_bkg" }
      ],
      "aux_obs": [
         { "name": "aux_bkg", "min_value": -5, "max_value": 5 }
      ],
      "channels": [
         {
            "name": "measurement_region",
            "type": "bin",
            "samples": [
               {
                  "name": "Signal",
                  "norm": "xs_signal",
                  "nominal_yields": [ 1 ]
               },
               {
                  "name": "Background",
                  "nominal_yields": [ 1 ],
                  "impacts": {
                     "np_bkg": 0.25
                  }
               }
            ]
         }
      ]
   },

   "data": {
      "channels": [
        { "name" : "measurement_region", "counts": 2 }
      ],
      "aux_obs": [
         { "name": "aux_bkg", "value": 0 }
      ]
   }
}
    \end{lstlisting}
    \caption{\label{fig:simple_model:listing} Specification for the example SLLS model described in the text.}
\end{figure}

The results for the signal yield $s$ are computed using its maximum likelihood estimator (MLE) $\hat{s}$ and the profile-likelihood ratio
\begin{equation}
    \Lambda(s) = -2\log\frac{L(s, \hat{\hat{b}}(s))}{L(\hat{s}, \hat{b})}
\end{equation}
where $L(s,b)$ is the likelihood function of the measurement, $\hat{b}$ is the MLE of $b$ and $\hat{\hat{b}}(s)$ its conditional MLE at a fixed value $s$ of the signal yield. The values of $\Lambda(s)$ can then be used to derive results such as confidence intervals on $s$ or the discovery significance of the signal.

\begin{figure}[tbp]
\centering
\begin{subfigure}[b]{0.48\textwidth}
    \includegraphics[width=\textwidth]{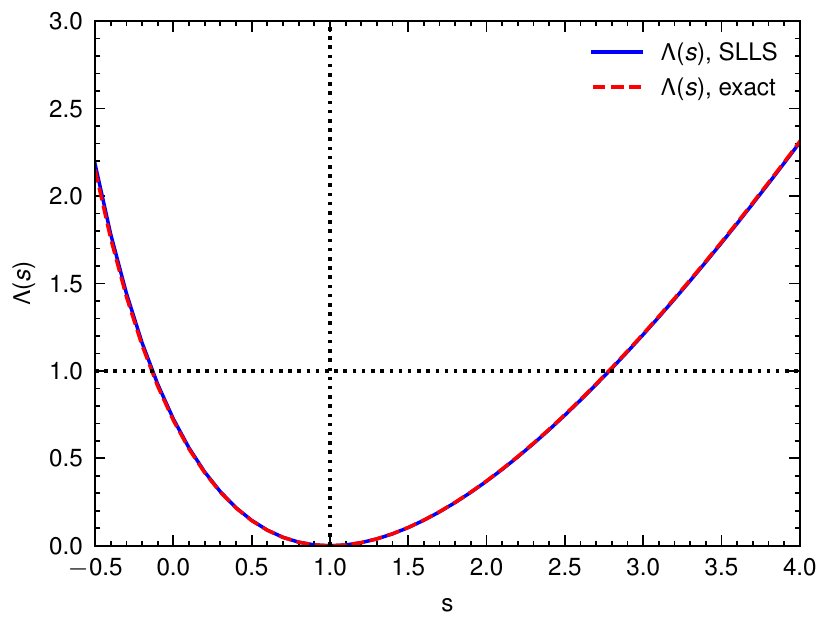}
    \caption{\label{fig:simple_model:scan}}
\end{subfigure}
\begin{subfigure}[b]{0.48\textwidth}
    \includegraphics[width=\textwidth]{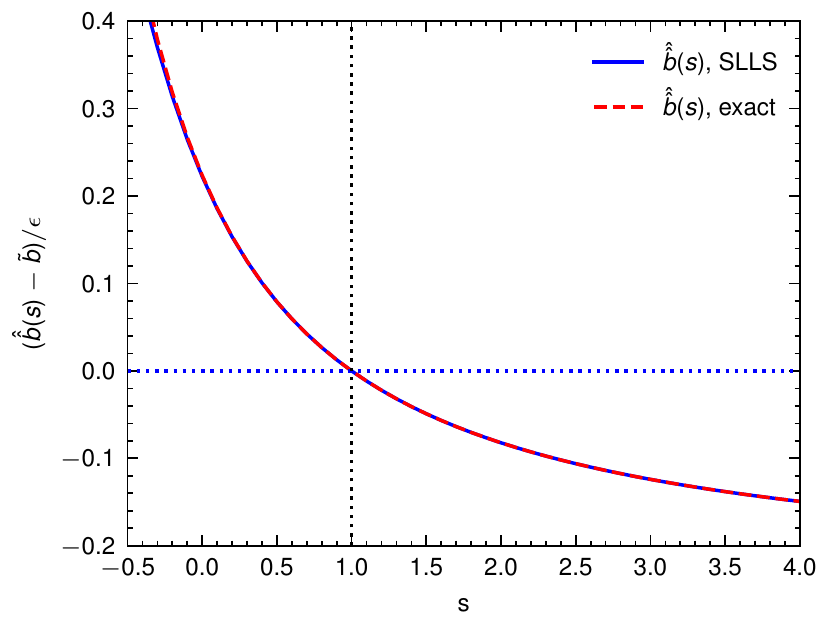}
    \caption{\label{fig:simple_model:spag}}
\end{subfigure}
\caption{\label{fig:simple_model:results} Values of (a) $\Lambda(s)$ and (b) the conditional MLE $\hat{\hat{b}}(s)$ for a range of values of the signal yield $s$, computed from the model described in the text. In each plot, the simplified likelihood result (solid blue) is compared to an exact closed-form expression of the same quantity (dashed red), showing very close agreement.}
\end{figure}

Figure~\ref{fig:simple_model:results} shows the values of $\Lambda(s)$ and $\hat{\hat{b}}(s)$ computed from the model given in Figure~\ref{fig:simple_model:listing} for a range of values of $s$. In this simple case both results can also be computed in closed form as
\begin{align}
\hat{\hat{b}}(s) &= \frac{1}{2} \left[\sqrt{(s + \tilde{b} - \tilde{b}^2\epsilon^2)^2 + 4 \tilde{b}^2\epsilon^2 n} - (s - \tilde{b} + \tilde{b}^2\epsilon^2)\right] \\
\Lambda(s) &= 2(s - \hat{s} + \hat{\hat{b}}(s) - \hat{b}) - 2 n\log \left(\frac{s + \hat{\hat{b}}(s)}{\hat{s} + \hat{b}}\right).
\label{eq:simple_pred}
\end{align}
and excellent agreement is observed between these expressions and the SLLS results.
The asymmetric shape of $\Lambda(s)$ is driven by the Poisson nature of the measurement, and the good agreement in this case is due to the fact that this feature is accounted for exactly in the simplified likelihood. Using a Gaussian approximation would yield a parabolic shape for $\Lambda(s)$ that would provide a less accurate description. While the systematic uncertainty on the background yield plays only a small role in this example, the good agreement in the profiled values $\hat{\hat{b}}(s)$ of the corresponding NP shows that systematic effects are also described accurately within the linear approximation.

This agreement is not by construction, since SLLS only provides an approximation to the exact results of Eq.~\ref{eq:simple_pred}. For instance deviations of about 10\% in the value of $\hat{\hat{b}}(s)$ and $\Lambda(s)$ are observed in a scenario in which the auxiliary background observable $\tilde{b}$ is set to deviate from the nominal value $b_0$ by $2\sigma$ in the lower direction, which in turn pulls $\hat{\hat{b}}(s)$ away from its nominal value. For the same reason, the SLLS computation of the total expected yield $s + \hat{\hat{b}}(s)$ can take negative values when using the linear expression of Eq.~\ref{eq:nexp_linear}, although this quantity is positive by construction in the exact computation. Since $\Lambda(s)$ cannot be computed for null or negative values of $s + \hat{\hat{b}}(s)$, this motivates the use of Eq.~\ref{eq:nexp_non_linear} which ensures positive-definite values for the expected event yields.

\section{Application to an ATLAS search for new phenomena}
\label{sec:trileptons}

\subsection{Full statistical model}

This section presents a realistic application of the SLLS framework to a search for new phenomena by the ATLAS collaboration~\cite{trileptons} for which the full experimental statistical model has been published~\cite{hepdata.99806.v2/r2}. The search targets supersymmetric particles in final states with at least three charged leptons originating from the chargino decay $\tilde{\chi}^+_1 \rightarrow Z\ell \rightarrow 3\ell$. The analysis considers three signal regions (SRs), targeting signatures with 3 leptons ($3\ell$), 4 leptons ($4\ell$) and 4 leptons with a fully reconstructed $W$, $Z$ or $H$ boson (FR). Each signal region is divided into 16 bins of the invariant mass $m_{Z\ell}$ of the trilepton system. Three single-bin control regions are also included to provide data-driven estimates of the main backgrounds, from the Standard Model production of a $WZ$ boson pair, a $ZZ$ pair, or a $t\bar{t}$ pair accompanied by a $Z$ boson ($t\bar{t}Z$). The model includes a single parameter of interest, the signal strength $\mu_{\text{signal}}$, and 624 NPs: three unconstrained parameters representing normalization terms for the main backgrounds and 621 constrained parameters representing systematic uncertainties.

The full statistical model of the analysis was published by the ATLAS collaboration as a \pyhf\ model available in the HEPData repository~\cite{hepdata.99806.v2/r2}. In this example we consider the example case of a chargino with a mass of $500\gev$ with branching ratios to $W$, $Z$ and $H$ bosons of respectively $20\%$, $60\%$ and $20\%$, and equal branching ratios to $e$, $\mu$ and $\tau$ for the accompanying lepton.

\subsection{Simplified model}

\begin{figure}[tbp]
\centering 
\begin{subfigure}[b]{0.45\textwidth}
    \includegraphics[width=\textwidth]{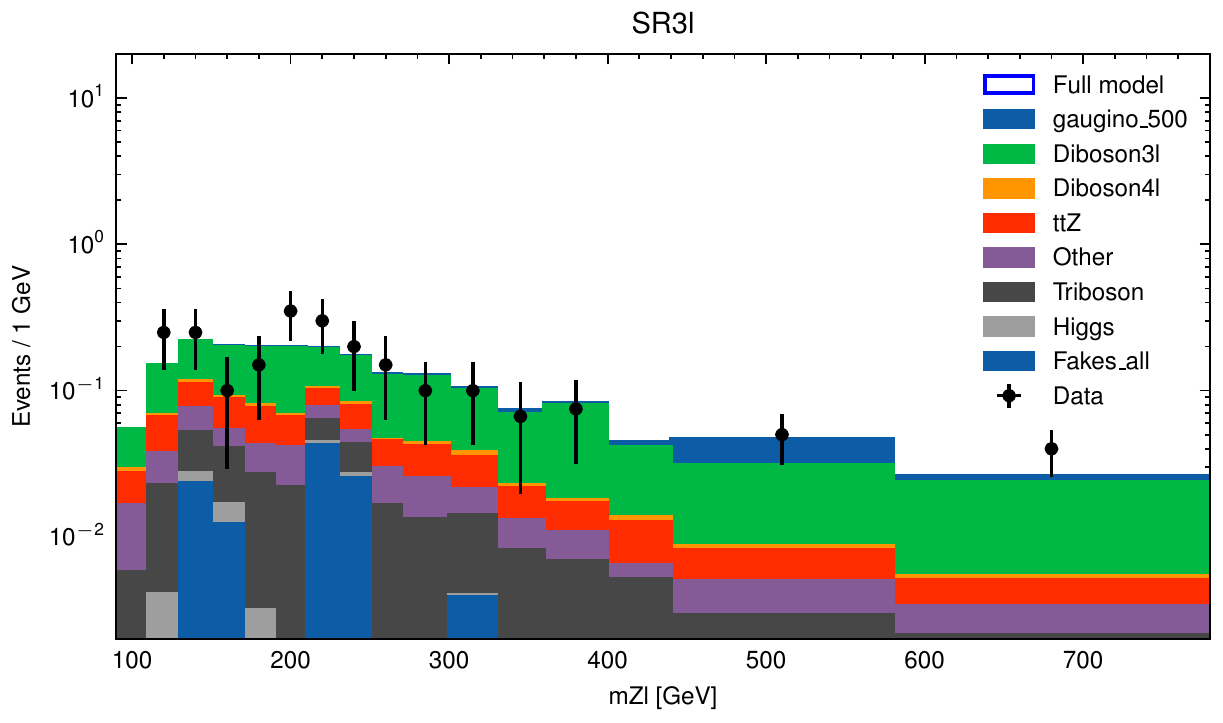}
    \caption{\label{fig:trileptons:SR3l}}
\end{subfigure}
\begin{subfigure}[b]{0.45\textwidth}
    \includegraphics[width=\textwidth]{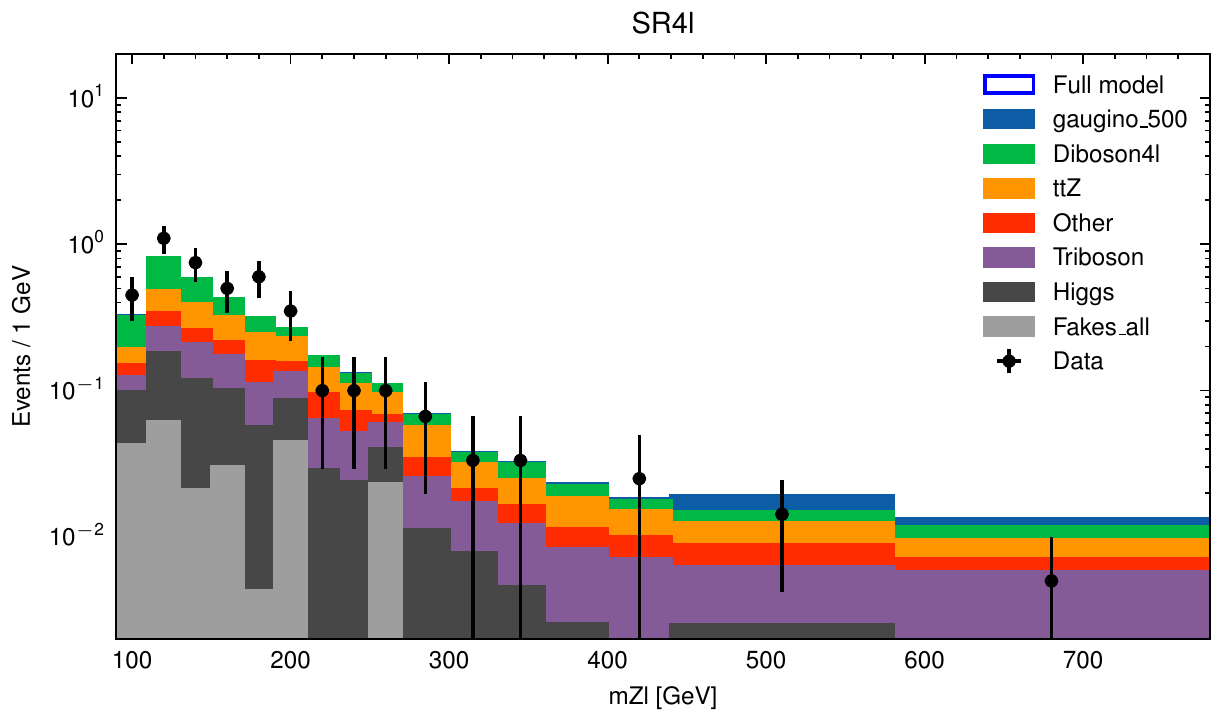}
    \caption{\label{fig:trileptons:SR4l}}
\end{subfigure}\\
\vspace{0.5cm}
\begin{subfigure}[t]{0.45\textwidth}
    \includegraphics[width=\textwidth]{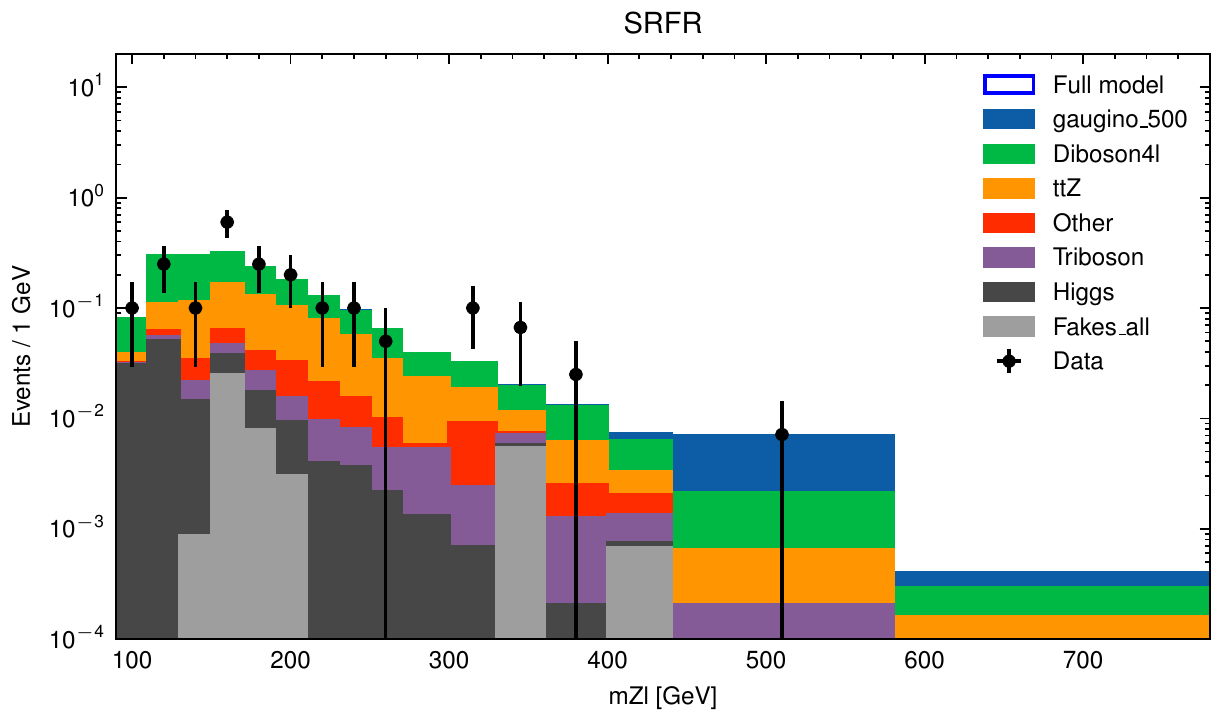}
    \caption{\label{fig:trileptons:SRFR}}
\end{subfigure}
\begin{subfigure}[t]{0.45\textwidth}
\vspace{-4cm}
    \includegraphics[width=\textwidth]{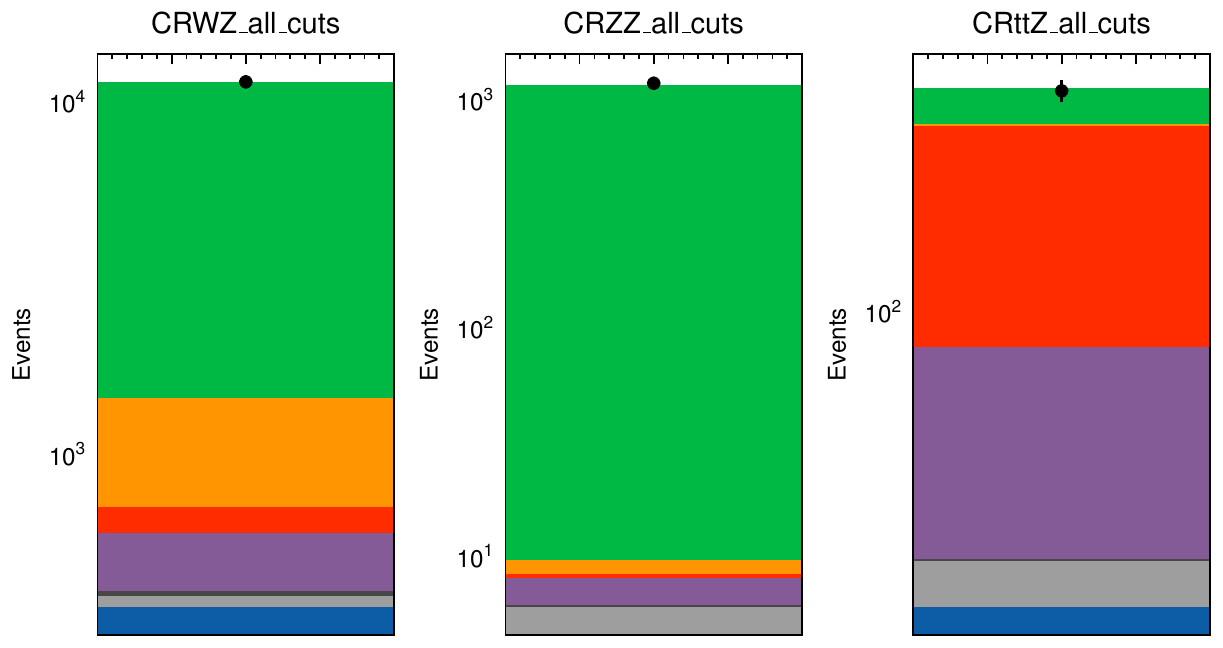}
\vspace{-0.2cm}
    \caption{\label{fig:trileptons:CRs}}
\end{subfigure}
\caption{\label{fig:trileptons:spectra} Expected and observed event counts in the SR3l, SR4l and SRFR signal regions of the analysis of Ref~\cite{trileptons}, shown respectively in panels (a), (b) and (c). Panel (d) shows the analysis control regions. The signal regions are binned in the $m_{Zl}$ observable, while the control regions each use a single inclusive event count. The observed data (black points) is overlaid with stacked histograms (filled areas) representing the gaugino signal (dark blue) and the main background contributions.}
\end{figure}

The SLLS model is computed by taking the nominal event yield for each signal and background sample in each bin of each region from the \pyhf\ model, as published by the ATLAS collaboration.\footnote{The nominal bin yields can be freely chosen, but it is clearly desirable to pick values close to the data best-fits since the linear approximation is more accurate for small deviations of the NPs from the nominal. One could also have chosen the yields obtained in a fit of the model to the observed data, which would have yielded a better linear approximation and possibly better agreement with the \pyhf\ model.} The $1\sigma$ impacts of the systematics NPs are similarly determined from the definition of the systematic effects in the \pyhf\ model. The impact of the background normalization NPs are derived from the relative fractions of the corresponding backgrounds. The conversion is performed using an automated tool included in the \fastprof\ package. The measurement regions of the analysis as implemented in the simplified model are shown in Figure~\ref{fig:trileptons:spectra}. 

\begin{figure}[tbp]
\centering 
\begin{subfigure}[b]{0.60\textwidth}
    \includegraphics[width=\textwidth]{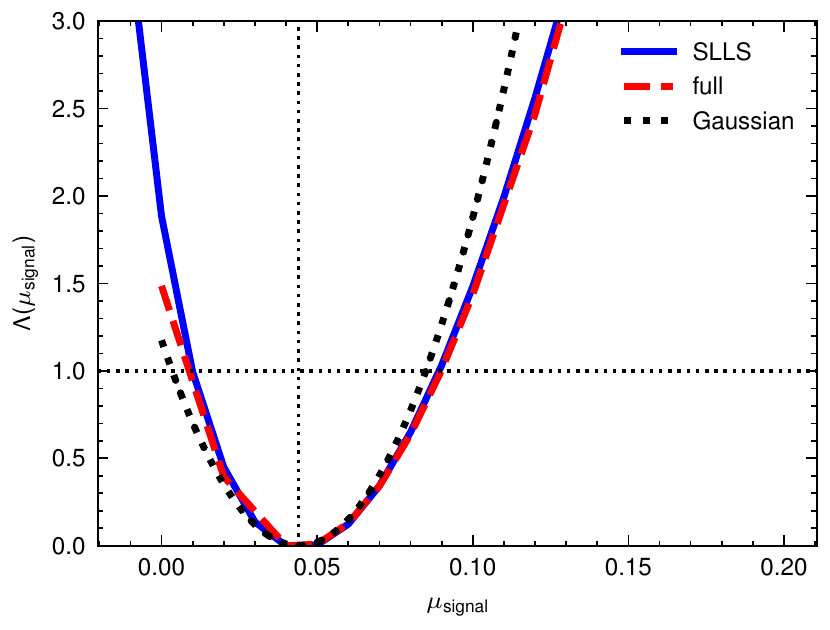}
    \caption{\label{fig:trileptons:scan}}
\end{subfigure}\\
\vspace{0.5cm}
\begin{subfigure}[b]{0.48\textwidth}
    \includegraphics[width=\textwidth]{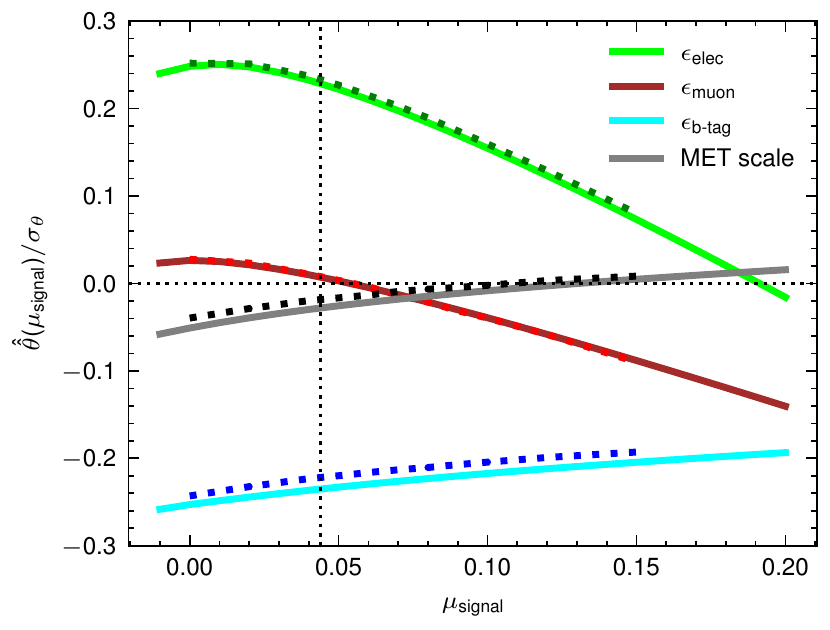}
    \caption{\label{fig:trileptons:nuis}}
\end{subfigure}
\begin{subfigure}[b]{0.48\textwidth}
    \includegraphics[width=\textwidth]{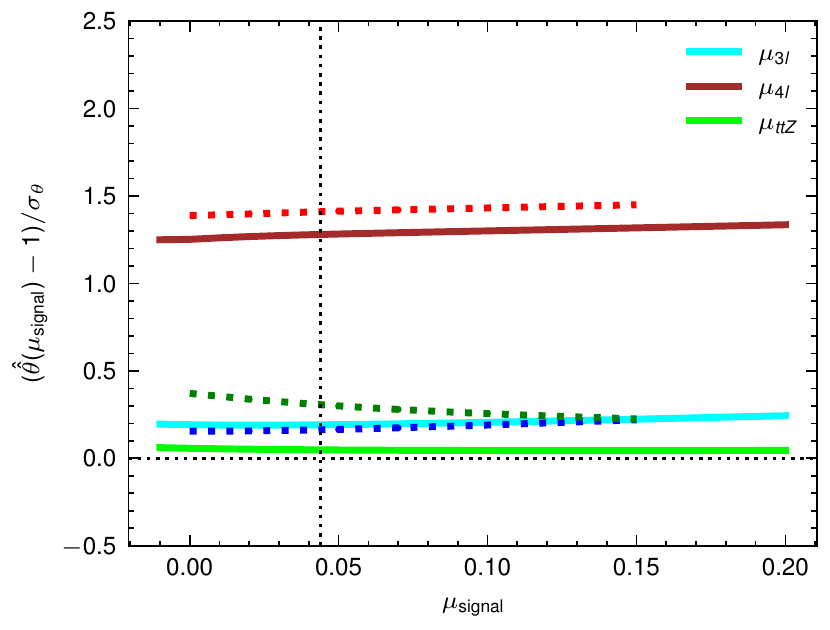}
    \caption{\label{fig:trileptons:kfac}}
\end{subfigure}
\caption{\label{fig:trileptons:results} Comparison between the SLLS simplified model (solid lines) and the full model (dashed lines) for (a) the PLR $\Lambda(\mu_{\text{signal}})$ as a function of $\mu_{\text{signal}}$, (b) the profiled values of selected NPs describing systematic uncertainties and (c) the profiled values of NPs describing scale factors applied to the normalization of the main analysis backgrounds. The profiled values are shown as deviations from the nominal value of the parameters (0 for systematic uncertainties, $1$ for background scaling factors), divided by the uncertainty on the parameter in the full-model fit to the observed data with free $\mu_{\text{signal}}$. The SLLS results are computed using the \fastprof\ tool, and the full-model results with the \pyhf\ tool. Panel (a) also shows the PLR scan computed using a Gaussian model as described in the text.}
\end{figure}

The profile likelihood scan of the signal strength parameter $\mu_{\text{signal}}$ using the simplified model is shown in Figure~\ref{fig:trileptons:scan}. A reference scan computed using the full model is also presented for comparison, and shows that the simplified likelihood provides an adequate description of the full result. A simple Gaussian model, using the best-fit value $\mu_{\text{signal}}$ in the observed data computed by \pyhf\ and the corresponding parabolic error, is also displayed and shows worse agreement. The 95\% CL$_s$ upper limit on $\mu_{\text{signal}}$ computed using the simplified model is 0.126, in good agreement with the value of 0.124 obtained using the full model. The Gaussian model yields a value of 0.114.

The fits to the SLLS likelihood with fixed $\mu_{\text{signal}}$ take about $50 \,\text{ms}$ on a laptop computer equipped with a 16-core Intel i7-10875H CPU. The fit with free $\mu_{\text{signal}}$, which relies on non-linear rather than linear minimization for this parameter (since POIs are treated exactly), takes about $0.5 \,\text{s}$. A full-likelihood fit performed with \pyhf\ require approximately $10 \,\text{min}$ on the same computing platform, a factor $\approx 1000$ longer. The full-likelihood fit times for the fixed-POI and free-POI cases are similar, since both are dominated by the non-linear minimization over the 624 NPs. These fits are performed with \numpy\ as the numerical backend to \pyhf, the same as used the \fastprof\ implementation of SLLS likelihoods. Better performance can however likely be achieved using other \pyhf\ backends interfacing to \tensorflow~\cite{tensorflow} or \pytorch~\cite{pytorch}.

\begin{figure}[tbp]
\centering 
\includegraphics[width=.60\textwidth,valign=t]{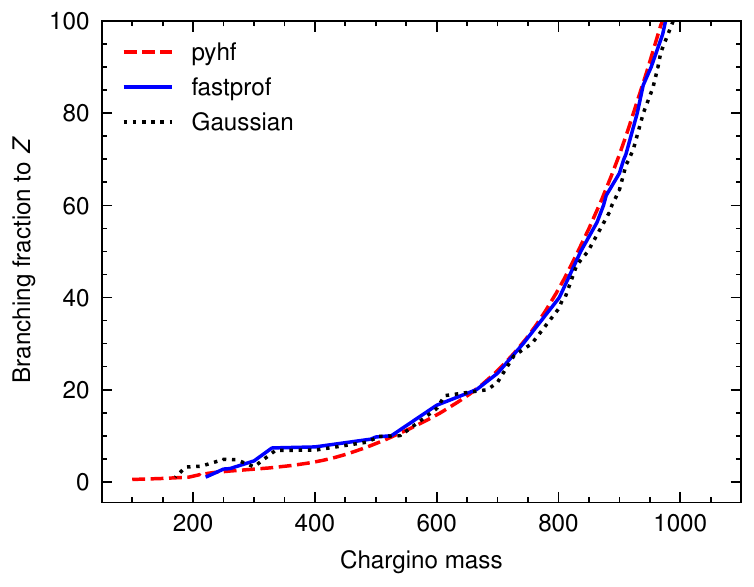}
\caption{\label{fig:trileptons:excl} Exclusion plot in the plane of chargino mass and its branching ratio to $Z$ bosons, assuming equal branching ratios to $W$ and $H$ and to all lepton flavors. The computation from SLLS simplified likelihoods (solid blue) is compared with a reference (dashed red) taken from the top-left panel of Figure 9 in Ref.~\cite{trileptons} and good agreement is observed. Gaussian models computed from the full likelihood as described in the text (dotted black) also show good agreement in this case.}
\end{figure}

To validate the SLLS linear profiling, the profiled values of selected NPs are shown in Figures~\ref{fig:trileptons:nuis} and~\ref{fig:trileptons:kfac} for both the SLLS and the full model. Good agreement is seen between the two cases, illustrating that the original likelihood function is modeled to good approximation at the level of individual NPs. The largest deviation is seen in the scale factor for the $t\bar{t}Z$ background, amounting to about 30\% of the fit uncertainty.
As a further illustration, the exclusion contour presented in Figure 9 of the original publication is recomputed using the SLLS models and compared to the full-model results. The results are shown in Figure~\ref{fig:trileptons:excl}, and good agreement is again observed. An exclusion contour based on Gaussian models built as described above is also presented for comparison and shows similar agreement in this case, in part due to the fact that the signal production cross-sections vary rapidly with the chargino mass.

\section{Simplified likelihoods for unbinned models}
\label{sec:unbinned}

\subsection{Binned description of unbinned models}

The previous examples use a binned description of the experimental measurement, which employs only two types of PDFs: Poisson distributions to describe the counting experiments in each bin, and Gaussian distributions for the constraints.
Another common modeling option is \emph{unbinned} models, which describe the continuous probability distribution of the measurement observables. It is used for instance to study the $H \rightarrow \gamma\gamma$ decay of the Higgs boson at the LHC~\cite{HggATLAS, HggCMS}, as well as in many results published by LHCb (see for instance Refs.~\cite{LHCb:2012skj, LHCb:2021vvq}). It requires support for arbitrary PDF forms, as needed to describe each measurement, and therefore more general and flexible tools than for binned models. For LHC measurements, this functionality is usually provided by the \roofit\ package~\cite{roofit} distributed as part of \ROOT, but this and other similar tools are not widely used outside the high-energy physics experimental community. While there are some recent ongoing efforts to provide more portable alternatives, none is currently in wide use.

A possible way forward is based on the observation that unbinned models can be approximated by binned models with a sufficiently fine binning (see Appendix~\ref{app:unbinned_approx}). While this approach typically runs into practical difficulties for full likelihoods due to the large number of bins required, it is feasible for simplified likelihoods which are quick to evaluate even for relatively large bin numbers. In the rest of this section, we present the application of the SLLS framework to an unbinned model loosely inspired by an ATLAS $H \rightarrow \gamma\gamma$ measurement.

\subsection{Full model example}

We consider a simple example based on the ATLAS $H \rightarrow \gamma\gamma$ analysis of Ref.~\cite{HggATLAS}. The analysis uses an unbinned model based on the distribution of the invariant mass $m_{\gamma\gamma}$ of the two photons in the range $105 < m_{\gamma\gamma} < 160\,\text{GeV}$. The Higgs boson signal manifests itself as a sharp peak in the $m_{\gamma\gamma}$ distribution, with a position close to the Higgs boson mass and a width of 1.1--2.1 GeV depending on event kinematics. The background contributions follow smoothly falling shapes. Several signal regions (referred to as \emph{categories} in the rest of this section) are defined according to the properties of the signal photons and of the rest of the event.

The example uses a simplified description of the 33 categories defined in Ref.~\cite{HggATLAS} to study Higgs boson production in the gluon-fusion process. The signal and background distributions are represented respectively by Gaussian and exponential distributions, instead of the more complex shapes used in Ref.~\cite{HggATLAS}. The peak position and width of the Gaussian, as well as the expected signal and background yields are taken from Ref.~\cite{HggATLAS}, while the exponential slope of the background is assumed to be $-0.02\, \text{GeV}^{-1}$ in all categories. The background normalizations and exponential slopes are free to vary in the fit, except for the slopes in five low-statistics categories which are kept fixed to avoid unstable fits.
Five NPs are used to describe the leading systematic uncertainties: the uncertainty on the integrated luminosity of the dataset; on the reference cross-section for the gluon-fusion production process; on the effect of parton shower modeling on the signal yields; on the \hgg\ reconstruction efficiency; and on the photon energy resolution. This last uncertainty leads to a change in the width of the signal peak and therefore induces highly non-linear effects in the per-bin signal yields in a binned description of the likelihood. Systematic uncertainties on the background model are implemented using separate NPs in each category, following the \emph{spurious signal} method described in Ref.~\cite{HggATLAS}. The values of the uncertainties listed above are all taken from Ref.~\cite{HggATLAS}. In total, 99 NPs are defined. The single POI is the Higgs boson signal strength $\mu$, applied as a scaling factor to the expected signal yield in all categories. A dataset of events randomly generated from the model PDF is used as the "observed" data in this example.

\subsection{Simplified model}

\begin{figure}[tbp]
\centering 
\includegraphics[width=.75\textwidth]{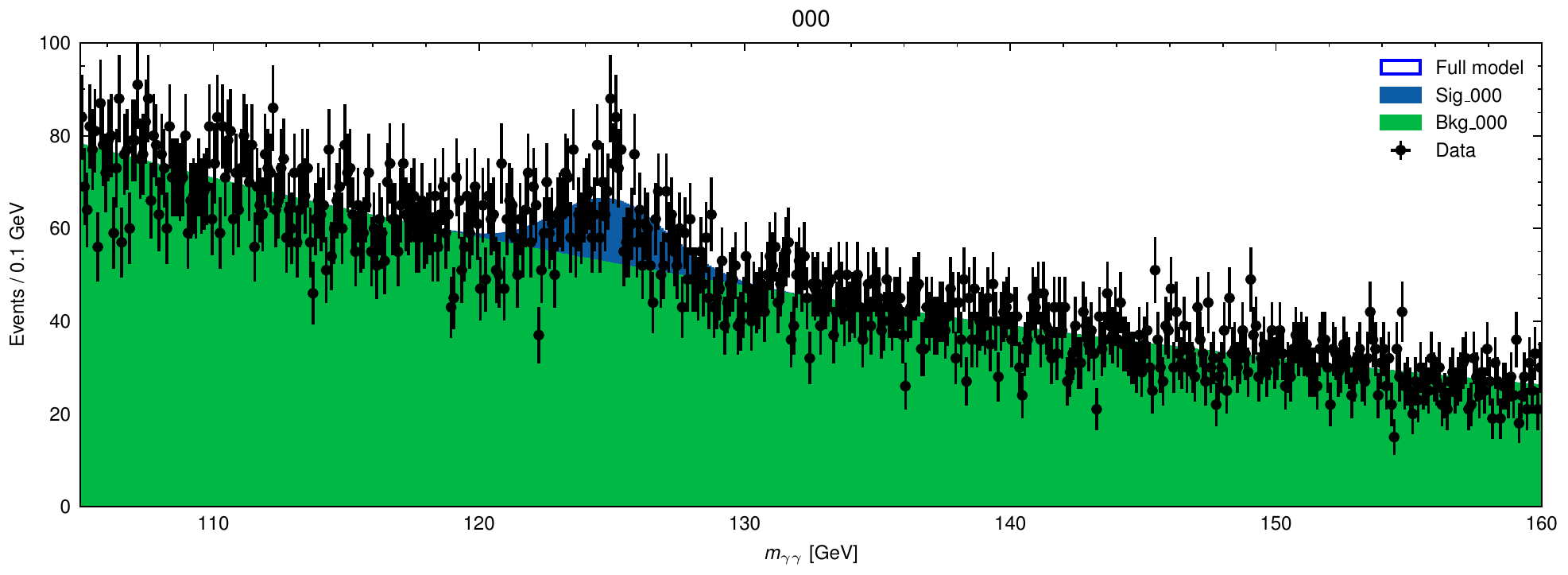} \\
\includegraphics[width=.75\textwidth]{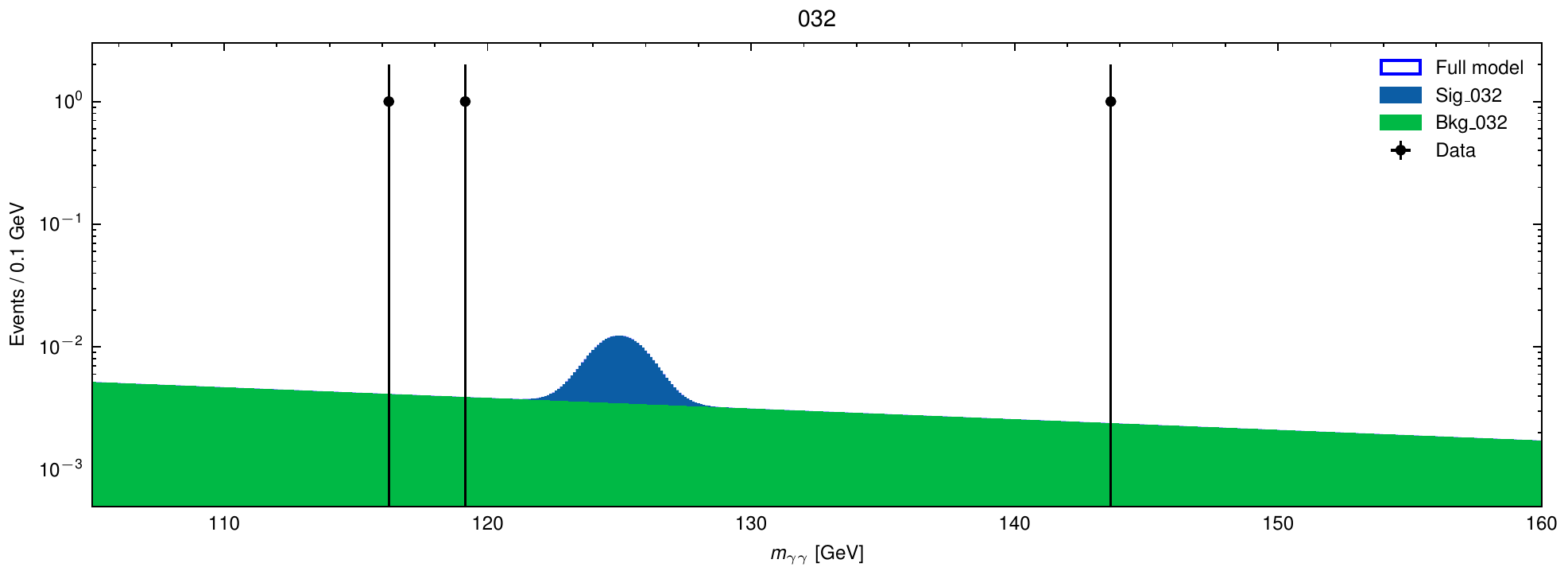}
\caption{\label{fig:unbinned:spec} Distributions of the observable $m_{\gamma\gamma}$ for two selected categories -- the ones labeled 0-jet, $p_\text{T}^H < 10\,\text{GeV}$ (top) and $p_\text{T}^H \ge 650\,\text{GeV}$ (bottom) in Ref.~\cite{HggATLAS}. The bin width is $0.1\,\text{GeV}$ in both cases. The signal and background contributions (blue and green histograms respectively) are shown together with the example dataset (black points), for the best-fit value of the model parameters.}
\end{figure}

The SLLS model is built as a binned approximation to the full model. A fine binning is required to obtain an accurate description of the signal peak. In this example a uniform bin width of $0.1\,\text{GeV}$ is used, leading to 18150 bins in total for the 33 categories\footnote{A variable-width binning with wider bins away from the peak can also be considered, but a uniform binning was chosen in this example for simplicity.}. The $m_{\gamma\gamma}$ distributions for two selected categories (the first and last in the order used in Ref.~\cite{HggATLAS}) are shown in Figure~\ref{fig:unbinned:spec}.

\begin{figure}[tbp]
\centering 
\begin{subfigure}[b]{0.48\textwidth}
    \includegraphics[width=\textwidth]{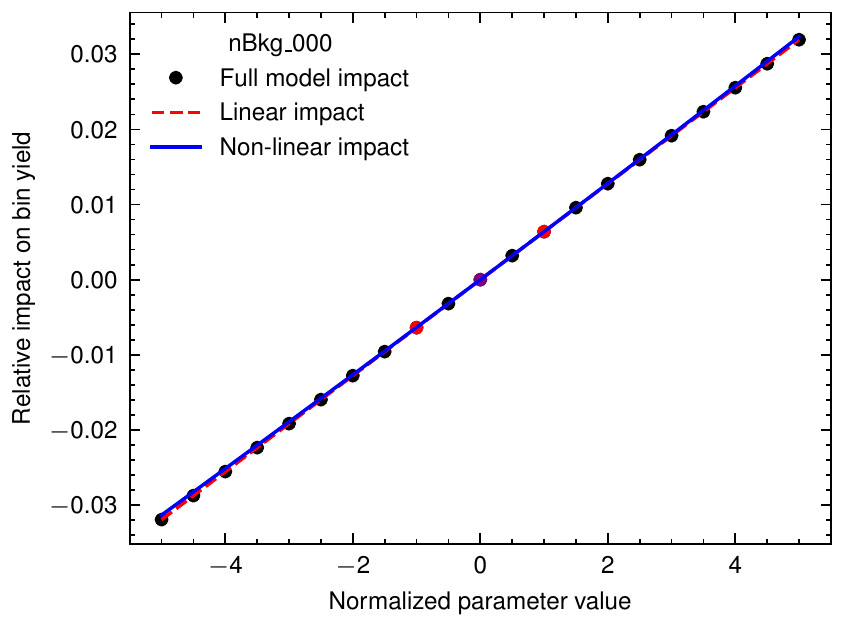}
    \caption{\label{fig:unbinned:nBkg}}
\end{subfigure}
\begin{subfigure}[b]{0.48\textwidth}
    \includegraphics[width=\textwidth]{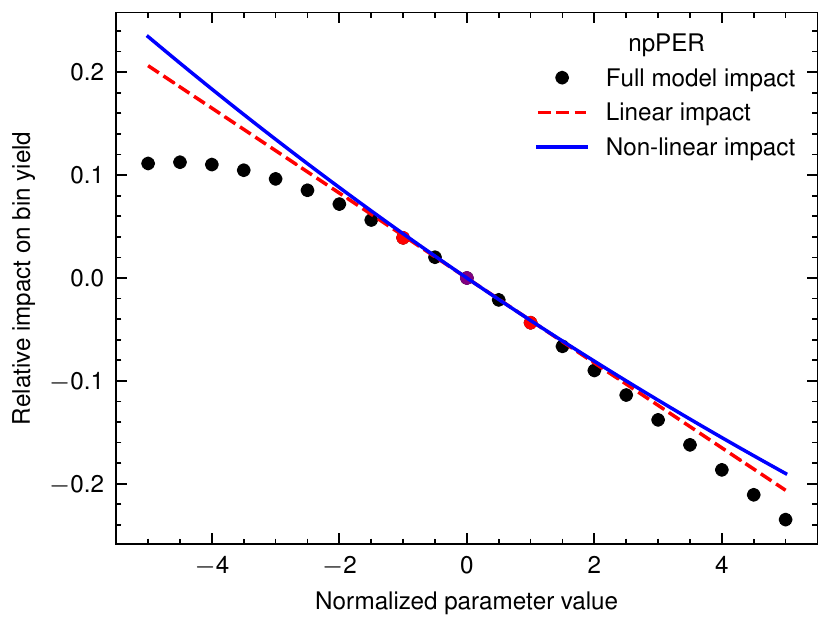}
    \caption{\label{fig:unbinned:PER}}
\end{subfigure}
\caption{\label{fig:unbinned:impacts} Relative change in expected bin yields as a function of the normalized parameter value for two cases: (a) the impact of the background normalization parameter \texttt{nBkg\_000} on the expected background yield; and (b) the impact of the photon energy resolution parameter \texttt{npPER} on the expected signal yield. In both cases, the bin belongs to the first category of the model and is located at $m_{\gamma\gamma} \approx 127\,\text{GeV}$, about $0.7\sigma$ above the signal peak. The impacts computed from the full model (dots) are compared with the linear impacts computed from Eq.~\ref{eq:nexp_linear} (dashed red line) and the non-linear impacts from Eq.~\ref{eq:nexp_non_linear} (solid blue line).}
\end{figure}

The conversion is performed using an automated tool distributed as part of the \fastprof\ package. All the NPs are retained, and their effect is described in terms of their linear impact on the event yield in each measurement bin, following the SLLS procedure. In some cases, in particular normalization parameters such as the one shown on Figure~\ref{fig:unbinned:nBkg}, impacts are linear by construction. By contrast, the photon energy resolution systematic shown in Figure~\ref{fig:unbinned:PER} exhibits non-linear behavior since it induces a change in the width of the signal peak which does not propagate linearly to the bin contents. Non-linearities become larger
closer to the tails of the signal peak, but with a smaller impact on the results due to lower signal yields. The linear approximation remains in any case typically accurate for small deviations of the NP from the nominal.
\begin{figure}[tbp]
\centering 
\begin{subfigure}[b]{0.60\textwidth}
    \includegraphics[width=\textwidth]{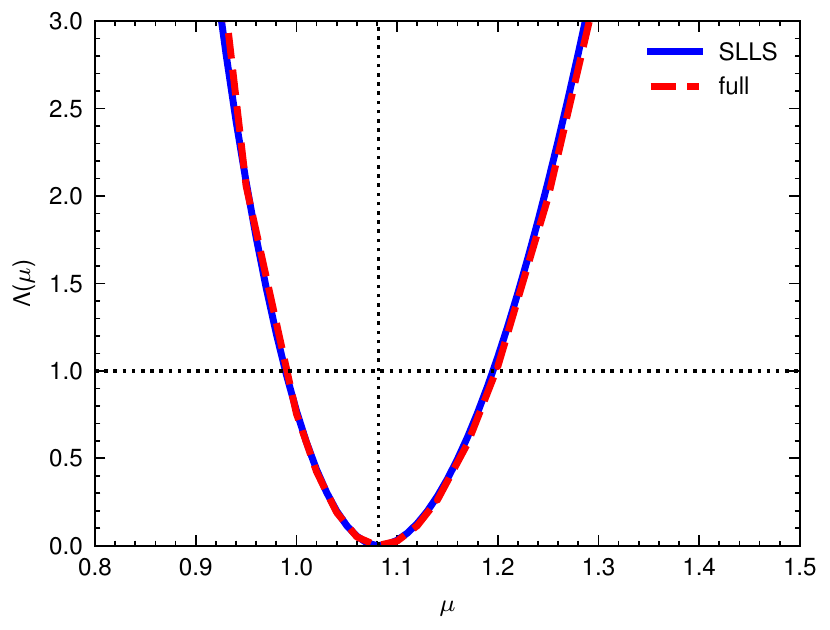}
    \caption{\label{fig:unbinned:scan}}
\end{subfigure}\\
\vspace{0.5cm}
\begin{subfigure}[b]{0.48\textwidth}
    \includegraphics[width=\textwidth]{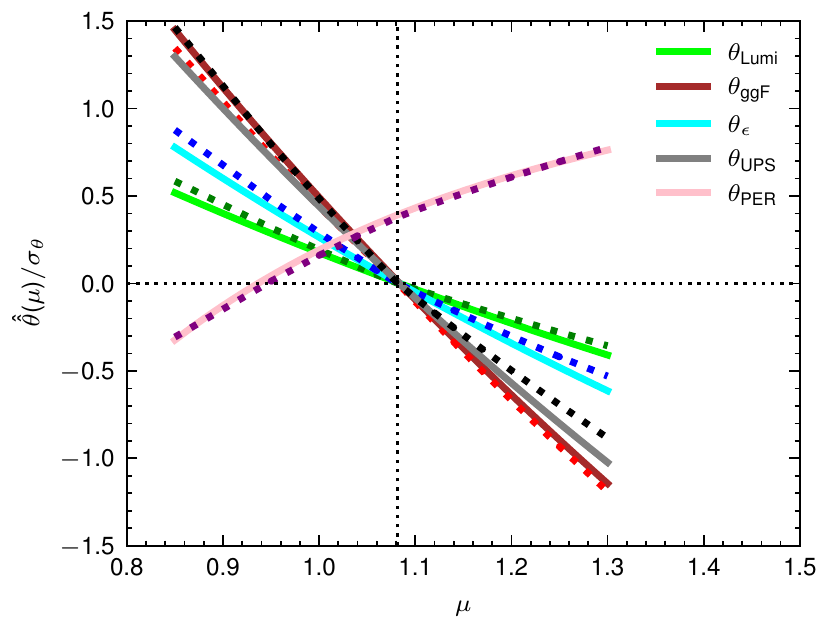}
    \caption{\label{fig:unbinned:nuis}}
\end{subfigure}
\begin{subfigure}[b]{0.48\textwidth}
    \includegraphics[width=\textwidth]{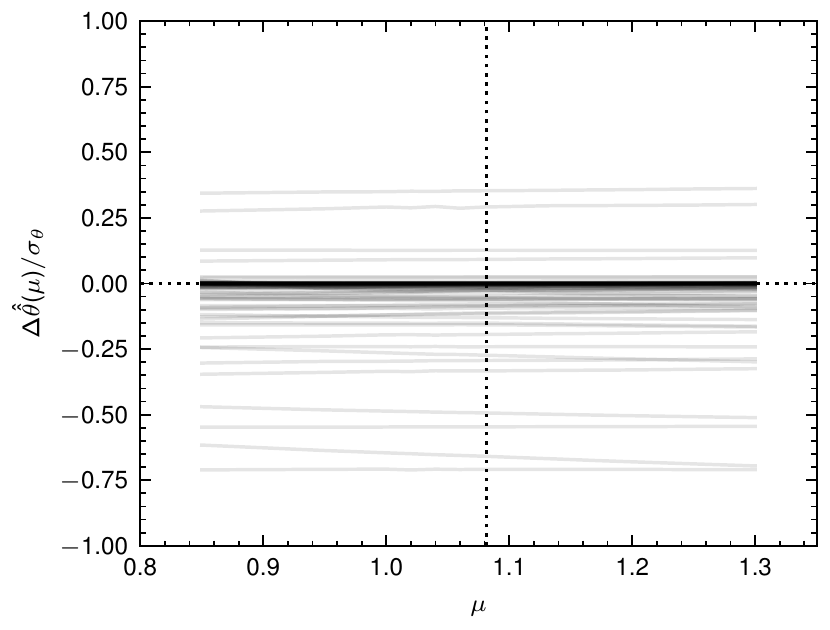}
    \caption{\label{fig:unbinned:spag}}
\end{subfigure}
\caption{\label{fig:unbinned:scans} Comparison between the SLLS model (solid lines) and the full model (dashed lines) for (a) the profile likelihood $\Lambda(\mu)$ as a function of the POI $\mu$; (b) the profiled values of a selection of NPs describing systematic uncertainties; and (c) the difference between the profiled values of each NP obtained from the SLLS model and the full model, divided by its uncertainty in the full-model fit to the data with free $\mu$.}
\end{figure}
Figure~\ref{fig:unbinned:scan} shows the profile likelihood scan for the signal strength parameter $\mu$ obtained with the linearized model. The reference result obtained with the full unbinned likelihood, computed using the \texttt{RooFitUtils} package\footnote{\url{https://gitlab.cern.ch/cburgard/RooFitUtils}}, is also shown for comparison and excellent agreement is observed. The resulting 68\% CL likelihood intervals are $\mu = 1.082^{+0.117}_{-0.093}$ for the full model and $\mu = 1.082^{+0.113}_{-0.093}$ for the simplified model. A fully Gaussian approximation, constructed as described in the previous section, yields $\mu = 1.082 \pm 0.098$. The fits take about $15\,\text{min}$ to perform on the full model, compared to about $50\,\text{ms}$ and $1\,\text{s}$ for simplified likelihood fits with respectively a free and floating $\mu$.

To better compare the treatment of systematic effects, the profiled values of the five NPs describing the leading systematic uncertainties are shown in Figure~\ref{fig:unbinned:nuis}. The agreement between the simplified and the full model is found to be accurate to about 10\% of the parameter uncertainties. This agreement is crucial to the description of this example since the uncertainty on $\mu$ is dominated by systematic effects. 
In particular, the profiling of the photon energy resolution systematic (which represents about 20\% of the total uncertainty) shows good agreement with the full model in spite of the non-linear effects highlighted in Figure~\ref{fig:unbinned:PER}.
Figure~\ref{fig:unbinned:spag} shows the difference between the profiled values of the other NPs in the simplified and full models, normalized to their fit uncertainty. This difference is below 10\% of the fit uncertainty for about 80\% of the parameters.

\section{Discussion}
\label{sec:discussion}

As observed in the examples shown in this paper, linearized NP impacts provide a generally adequate approximation of their behavior in the full model, in particular in the description of systematic effects.  It can be noted that the approximation approaches the exact description in situations where the impact of the NP is naturally linear, such as the case shown in Figure~\ref{fig:unbinned:nBkg}. Discrepancies are expected in cases which deviate from this ideal configuration, in particular for:
\begin{itemize}
    \item Large non-linear systematic uncertainties, with effects that are not fully accounted for in the linear approximation, such as the one shown in Figure~\ref{fig:unbinned:PER}.
    \item Asymmetric systematic uncertainties, with different impacts for NPs above or below their nominal value. These effects cannot be included in the linearized profiling, although they can be taken into account for the evaluation of the likelihood function.
    \item Low expected event yields, leading to Poisson counts that are not well-described by Gaussian distributions. While the Poisson distribution itself is described exactly in the SLLS formalism, non-linearities can occur due to systematics on the expected event yield, since the Poisson PDF does not depend linearly on its expected yield.
\end{itemize}
These situations are partially covered in the examples described in this paper, and it is encouraging that in these cases at least, the linear description seems adequate. However simplified likelihoods should be carefully validated against the full model in each case nevertheless. Tools to perform these checks are included in the \fastprof\ package, using methods similar to those shown in this paper.

Another limitation to take into consideration is the memory footprint of the $\Delta_{cbsk}$ coefficients which encode the linear impacts: their number is given by $N_{\text{NPs}} \times N_{\text{bins}} \times N_{\text{samples}}$, which can reach $O(10^8)$ or more for complex models. For models with a large numbers of bins and NPs, such as converted unbinned models, memory constraints can be more stringent than those related to computation times, since these computations mainly involves matrix operations that are quite efficient even for large models.

\section{Conclusion}
\label{sec:conclusion}

Simplified likelihoods provide a convenient setting for the reuse of experimental results, and functionality that is complementary to that of full models and Gaussian approximations.
The SLLS framework is based on a linear description of NP impacts which provides an approximation with two main benefits: it describes the Poisson behavior of counting measurements exactly, and also preserves the NPs of the full model and therefore a fully granular description of its systematic uncertainties. Both of these aspects make it well-suited to the description of LHC measurements, where systematic uncertainties and non-Gaussian effects from low event counts (e.g. in tails of distributions) both play important roles. 
The preservation of the NP structure allows in particular a proper treatment of correlated systematic effects when performing combinations of measurements, by identifying parameters associated with identical sources of uncertainty in the combination inputs.
Since the POIs of the original model are also preserved, reinterpretations of the simplified model can also be performed as for the original model. These properties have particular relevance to global combinations of LHC measurements, for instance those performed in the context of effective-field theory models~\cite{Ellis:2020unq,Hartland:2019bjb,Ethier:2021bye,Almeida:2021asy,Dawson:2020oco} which are based on measurements dominated by systematic uncertainties as well as others performed in high-momentum regions with low expected event counts. Currently these combinations are typically performed under Gaussian approximations without accounting for correlated uncertainties and Poisson behavior, and the use of simplified likelihoods could improve their accuracy.

SLLS models can be built automatically from binned likelihood implemented using the \HistFactory\ formalism within the \ROOT\ and \pyhf\ framework, or from unbinned likelihood using binned approximations. An implementation of the SLLS framework is provided in the \fastprof\ package at \url{https://github.com/fastprof-hep/fastprof}. The models are stored in a plain-text JSON format, and computations and other operations are performed using python tools based on the widely available \numpy\ and \scipy\ libraries. Together with other full and simplified likelihood formats with complementary functionality, it is hoped that the SLLS framework will encourage the further publication of detailed statistical models by LHC experiments and beyond.

\acknowledgments

The author would like to thank Nick Wardle providing the code for the simplified likelihoods of Ref.~\cite{Buckley:2018vdr}, and Tetiana Hryn'ova for valuable feedback.
Plots in this paper were produced with \texttt{matplotlib} using the \texttt{SciencePlots} style package~\cite{SciencePlots}.
This research was funded, in whole or in part, by l’Agence Nationale de la Recherche (ANR), project ANR-22-CE31-0022.

\clearpage

\appendix
\section{Linearization procedure}
\label{app:linearization}

This section provides a sketch of the derivation of the profile value $\hat{\hat{\theta}}_k(\vm)$ of the NP $\theta_k$ given by Eq~\ref{eq:profile}. Starting from the likelihood in Eq.~\ref{eq:L} and applying the linearization procedure described in Section~\ref{sec:linearization}, we obtain the negative log-likelihood
\begin{equation}
    \lambda(\vm, \vt) = 
    \sum\limits_{c=1}^{N_{\text{channels}}}
    \sum\limits_{b=1}^{N_{\text{bins}, c}}
    \left[\nu_{cb}(\vm, \vt) - n_{cb} \log \nu_{cb}(\vm, \vt)  \right]
    + \frac{1}{2} (\vt - \vtt)\Gamma(\vt - \vtt)
\end{equation}
up to a an additive constant. In the expression above, indices $c$, $b$ and $s$ run respectively over measurement channels, bins within each channel, and event samples. The $\nu_{cb}(\vm, \vt) = \sum_s \nu_{cbs}(\vm, \vt)$ are the total expected events yields for the corresponding channel bin, and the per-sample yields $\nu_{cbs}(\vm, \vt)$ are given by Eq~\ref{eq:nexp_linear}. The Gaussian constraints on the NPs are parameterized using the auxiliary observables $\vtt$ and the inverse covariance matrix $\Gamma$.

The derivative of $\lambda(\vm, \vt)$ with respect to the NPs \vt\ is
\begin{equation}
\frac{\partial \lambda}{\partial \vt} (\vm, \vt) =    \sum\limits_{c=1}^{N_{\text{channels}}}
    \sum\limits_{b=1}^{N_{\text{bins}, c}}
    \left[\sum\limits_s \nu_{cbs}\nom(\vm) \vD_{cbs} \left(1 - \frac{n_{cb}}{\nu_{cb}(\vm, \vt)}\right) \right]
    + \Gamma(\vt - \vtt).
\end{equation}
where $\vD_{cbs}$ is the vector with components $\Delta_{cbsk}$, the linear impacts of the parameter $\theta_k$ on $\nu_{cbs}$.
The linear approximation of NP impacts is applied to the denominator as
\begin{equation}
\frac{n_{cb}}{\nu_{cb}(\vm, \vt)}
\approx \frac{n_{cb}}{\nu\nom_{cb}(\vm)} \left[1 - \sum\limits_s\frac{\nu\nom_{cbs}(\vm)}{\nu\nom_{cb}(\vm)}\vD_{cbs} (\vt - \vt\nom) \right].
\end{equation}
and one finally obtains
\begin{equation}
\frac{\partial \lambda}{\partial \vt} (\vm, \vt) =  
Q(\vm) + P(\vm) \left[ \vt - \vt\nom\right] + \Gamma \left[ \vt - \vtt\right]
\end{equation}
with $Q(\vm)$ and $P(\vm)$ defined by Eq.~\ref{eq:pq}, and the profile values $\hat{\that}(\vm)$ defined by $\partial \lambda/\partial \vt (\vm, \hat{\that}(\vm)) = 0$ are therefore given by Eq.~\ref{eq:profile}.

\section{Binned approximation to an unbinned PDF}
\label{app:unbinned_approx}

We consider an extended unbinned PDF for an observable $x$,

\begin{equation}
    P(\vx; \theta)\prod\limits_{i=1}^n dx_i  = \frac{e^{-N(\vt)}}{n!} N(\vt)^n \prod\limits_{i=1}^n f(x_i; \vt) dx_i
    \label{eq:unbinned_L}
\end{equation}
where $f(x, \vt)$ is the PDF for one observation of $x$, the dataset consists of the values $x_1 \cdots x_n$, $\vt$ are the model parameters, and the expected number of observations is $N(\vt)$. We include the infinitesimal volume elements $dx_i$ in the expression since these will be useful below.

We introduce a set of bins $B_a$, $a = 1 \cdots N_{\text{bins}}$ that span the allowed range of $x$. In the spirit of finite-element analysis, we approximate $f(x)$ by a form that is constant over each bin as
\begin{subequations}
\begin{equation}
    f(x, \vt) = \sum\limits_i f_a(\vt) I_a(x)
\end{equation}
\begin{equation}
    f_a(\vt) = \frac{1}{w_a} \int_{B_a} f(x, \vt) dx
\end{equation}
where the indicator $I_a(x)$ is $1$ if $x \in B_a$ and $0$ otherwise, and $w_a = \int_{B_a} I_a(x) dx$ is the measure of bin $B_a$.

The value $f_a(\vt)$ is the average of $f(x, \vt)$ over the bin $B_a$, so that for a sufficiently fine binning and smooth $f(x, \vt)$,  $f(x, \vt) \approx f_a(\vt)$ for $x \in B_a$.

One can remove the explicit dependence on the $x_i$ by integrating them out of the likelihood. The integration of the product term of Eq.~\ref{eq:unbinned_L} can be written as
\begin{equation}
    \int \prod\limits_{i=1}^n f(x_i, \vt) dx_i = 
    \prod\limits_{i=1}^n \sum\limits_i f_a(\vt)\int I_a(x_i) dx_i = 
    \prod\limits_{i=1}^n f_{a_i}(\vt)w_{a_i} =
    \prod\limits_{a=1}^{N_{\text{bins}}} \left[f_a(\vt)w_a \right]^{n_a}
\end{equation}
where $a_i$ is the index of the bin to which $x_i$ belongs, $n_a$ is the number of observations that fall in bin $B_a$, and we have used the fact that the $x_i$ are independent to propagate the integral through the product. Returning to the full expression of Eq.~\ref{eq:unbinned_L}, we can write the likelihood as a function of the \vn\ as
\begin{equation}
    P(\vn; \theta) = \frac{e^{-N(\vt)}}{n_1! \cdots n_{N_{\text{bins}}}!} N(\vt)^n \prod\limits_{a=1}^{N_{\text{bins}}} \left[ w_a f_a(\vt) \right]^{n_a},
\end{equation}
 after including an additional multiplicative factor $(n, n_1, n_2, \cdots n_{N_{\text{bins}}})$ to account for the number of different orderings of the $x_i$ that can yield a given set of $n_a$.
One can introduce the per-bin expected yields
\begin{equation}
N_a(\vt) = w_a f_a(\vt) N(\vt)
\label{eq:bin_exp}
\end{equation}
and note that since
\begin{equation*}
    1 = \int f(x, \vt) dx = \sum\limits_{a=1}^{N_{\text{bins}}} \int_{B_a} f(x, \vt) dx = \sum\limits_{a=1}^{N_{\text{bins}}} w_a f_a(\vt)
\end{equation*}
 one has $N(\vt) = \sum_a N_a(\vt)$ as expected.
One can finally rewrite
\begin{equation}
    P(\vn; \theta) =  \prod\limits_{a=1}^{N_{\text{bins}}} \frac{e^{-N_a(\vt)}}{n_a!} N(\vt)^{n_a} \left[w_a f_a(\vt)\right]^{n_a} =  \prod\limits_{a=1}^{N_{\text{bins}}}  \frac{e^{-N_a(\vt)}}{n_a!} N_a(\vt)^{n_a}.
\end{equation}
This takes the usual form of a binned likelihood, with a Poisson distribution in each measurement bin with expected yields given by Eq.~\ref{eq:bin_exp}.
\end{subequations}








\bibliographystyle{JHEP}
\bibliography{fast_likelihoods}

\providecommand{\href}[2]{#2}\begingroup\raggedright\begin{thebibliography}{10}

\bibitem{Asimov}
G.~Cowan, K.~Cranmer, E.~Gross, and O.~Vitells, {\it Asymptotic formulae for
  likelihood-based tests of new physics},  {\em The European Physical Journal
  C} {\bf 71} (feb, 2011).

\bibitem{CMS:2022dwd}
{CMS Collaboration}, {\it {A portrait of the Higgs boson by the CMS experiment
  ten years after the discovery}},  {\em Nature} {\bf 607} (2022), no.~7917
  60--68, [\href{http://arxiv.org/abs/2207.00043}{{\tt arXiv:2207.00043}}].

\bibitem{ROOT}
R.~Brun and F.~Rademakers, {\it Root - an object oriented data analysis
  framework},  in {\em AIHENP'96 Workshop, Lausane}, vol.~389, pp.~81--86,
  1996.

\bibitem{FAIR}
M.~D. Wilkinson et~al., {\it The fair guiding principles for scientific data
  management and stewardship},  {\em Scientific data} {\bf 3} (2016).

\bibitem{RIF1}
W.~Abdallah et~al., {\it Reinterpretation of {LHC} results for new physics:
  Status and recommendations after run 2},  {\em {SciPost} Physics} {\bf 9}
  (aug, 2020).

\bibitem{RIF2}
K.~Cranmer et~al., {\it Publishing statistical models: Getting the most out of
  particle physics experiments},  {\em {SciPost} Physics} {\bf 12} (jan, 2022).

\bibitem{pyhf_joss}
L.~Heinrich, M.~Feickert, G.~Stark, and K.~Cranmer, {\it pyhf: pure-python
  implementation of histfactory statistical models},  {\em Journal of Open
  Source Software} {\bf 6} (2021), no.~58 2823.

\bibitem{pyhf_soft}
L.~Heinrich, M.~Feickert, and G.~Stark, {\it pyhf: v0.7.0}, .
  https://github.com/scikit-hep/pyhf/releases/tag/v0.7.0.

\bibitem{Buckley:2018vdr}
A.~Buckley, M.~Citron, S.~Fichet, S.~Kraml, W.~Waltenberger, and N.~Wardle,
  {\it {The Simplified Likelihood Framework}},  {\em JHEP} {\bf 04} (2019) 064,
  [\href{http://arxiv.org/abs/1809.05548}{{\tt arXiv:1809.05548}}].

\bibitem{Collaboration:2242860}
{CMS Collaboration}, {\it {Simplified likelihood for the re-interpretation of
  public CMS results}},  tech. rep., CERN, Geneva, 2017.

\bibitem{Coccaro:2019lgs}
A.~Coccaro, M.~Pierini, L.~Silvestrini, and R.~Torre, {\it {The DNNLikelihood:
  enhancing likelihood distribution with Deep Learning}},  {\em Eur. Phys. J.
  C} {\bf 80} (2020), no.~7 664, [\href{http://arxiv.org/abs/1911.03305}{{\tt
  arXiv:1911.03305}}].

\bibitem{Fichet:2016gvx}
S.~Fichet, {\it {Taming systematic uncertainties at the LHC with the central
  limit theorem}},  {\em Nucl. Phys. B} {\bf 911} (2016) 623--637,
  [\href{http://arxiv.org/abs/1603.03061}{{\tt arXiv:1603.03061}}].

\bibitem{Cranmer:2013hia}
K.~Cranmer, S.~Kreiss, D.~Lopez-Val, and T.~Plehn, {\it {Decoupling Theoretical
  Uncertainties from Measurements of the Higgs Boson}},  {\em Phys. Rev. D}
  {\bf 91} (2015), no.~5 054032, [\href{http://arxiv.org/abs/1401.0080}{{\tt
  arXiv:1401.0080}}].

\bibitem{HistFactory}
K.~Cranmer, G.~Lewis, L.~Moneta, A.~Shibata, and W.~Verkerke, {\it
  {HistFactory: A tool for creating statistical models for use with RooFit and
  RooStats}},  tech. rep., New York U., New York, 2012.
\newblock \url{https://cds.cern.ch/record/1456844}.

\bibitem{numpy}
C.~R. Harris et~al., {\it Array programming with {NumPy}},  {\em Nature} {\bf
  585} (Sept., 2020) 357--362.

\bibitem{scipy}
P.~Virtanen et~al., {\it {{SciPy} 1.0: Fundamental Algorithms for Scientific
  Computing in Python}},  {\em Nature Methods} {\bf 17} (2020) 261--272.

\bibitem{trileptons}
{ATLAS Collaboration}, {\it {Search for trilepton resonances from chargino and
  neutralino pair production in $\sqrt{s}$ = 13 TeV $pp$ collisions with the
  ATLAS detector}},  {\em Phys. Rev. D} {\bf 103} (2021), no.~11 112003,
  [\href{http://arxiv.org/abs/2011.10543}{{\tt arXiv:2011.10543}}].

\bibitem{hepdata.99806.v2/r2}
{ATLAS Collaboration}, {\it {Full likelihood of \em{Search for trilepton
  resonances from chargino and neutralino pair production in $\sqrt{s}$ = 13
  TeV $pp$ collisions with the ATLAS detector} (Version 2)}},  2021.
\newblock \url{https://doi.org/10.17182/hepdata.99806.v2/r2}.

\bibitem{tensorflow}
M.~Abadi et~al., {\it {TensorFlow}: Large-scale machine learning on
  heterogeneous systems},  2015.
\newblock Software available from tensorflow.org.

\bibitem{pytorch}
A.~Paszke et~al., {\it Pytorch: An imperative style, high-performance deep
  learning library},  in {\em Advances in Neural Information Processing Systems
  32}, pp.~8024--8035.
\newblock Curran Associates, Inc., 2019.

\bibitem{HggATLAS}
{ATLAS Collaboration}, {\it {Measurement of the properties of Higgs boson
  production at $\sqrt{s} = 13$ TeV in the $H\to\gamma\gamma$ channel using
  $139$ fb$^{-1}$ of $pp$ collision data with the ATLAS experiment}},  2022.
\newblock https://arxiv.org/abs/2207.00348.

\bibitem{HggCMS}
{CMS Colllaboration}, {\it {Measurements of Higgs boson properties in the
  diphoton decay channel in proton-proton collisions at $\sqrt{s} =$ 13 TeV}},
  {\em JHEP} {\bf 11} (2018) 185, [\href{http://arxiv.org/abs/1804.02716}{{\tt
  arXiv:1804.02716}}].

\bibitem{LHCb:2012skj}
{\bf LHCb} Collaboration, R.~Aaij et~al., {\it {First Evidence for the Decay
  $B_s^0 \to \mu^+ \mu^-$}},  {\em Phys. Rev. Lett.} {\bf 110} (2013), no.~2
  021801, [\href{http://arxiv.org/abs/1211.2674}{{\tt arXiv:1211.2674}}].

\bibitem{LHCb:2021vvq}
{\bf LHCb} Collaboration, R.~Aaij et~al., {\it {Observation of an exotic narrow
  doubly charmed tetraquark}},  {\em Nature Phys.} {\bf 18} (2022), no.~7
  751--754, [\href{http://arxiv.org/abs/2109.01038}{{\tt arXiv:2109.01038}}].

\bibitem{roofit}
W.~Verkerke and D.~Kirkby, {\it The roofit toolkit for data modeling},
  \href{http://arxiv.org/abs/physics/0306116}{{\tt physics/0306116}}.

\bibitem{Ellis:2020unq}
J.~Ellis, M.~Madigan, K.~Mimasu, V.~Sanz, and T.~You, {\it {Top, Higgs, Diboson
  and Electroweak Fit to the Standard Model Effective Field Theory}},  {\em
  JHEP} {\bf 04} (2021) 279, [\href{http://arxiv.org/abs/2012.02779}{{\tt
  arXiv:2012.02779}}].

\bibitem{Hartland:2019bjb}
N.~P. Hartland, F.~Maltoni, E.~R. Nocera, J.~Rojo, E.~Slade, E.~Vryonidou, and
  C.~Zhang, {\it {A Monte Carlo global analysis of the Standard Model Effective
  Field Theory: the top quark sector}},  {\em JHEP} {\bf 04} (2019) 100,
  [\href{http://arxiv.org/abs/1901.05965}{{\tt arXiv:1901.05965}}].

\bibitem{Ethier:2021bye}
{\bf SMEFiT} Collaboration, J.~J. Ethier, G.~Magni, F.~Maltoni, L.~Mantani,
  E.~R. Nocera, J.~Rojo, E.~Slade, E.~Vryonidou, and C.~Zhang, {\it {Combined
  SMEFT interpretation of Higgs, diboson, and top quark data from the LHC}},
  {\em JHEP} {\bf 11} (2021) 089, [\href{http://arxiv.org/abs/2105.00006}{{\tt
  arXiv:2105.00006}}].

\bibitem{Almeida:2021asy}
E.~d.~S. Almeida, A.~Alves, O.~J.~P. \'Eboli, and M.~C. Gonzalez-Garcia, {\it
  {Electroweak legacy of the LHC run II}},  {\em Phys. Rev. D} {\bf 105}
  (2022), no.~1 013006, [\href{http://arxiv.org/abs/2108.04828}{{\tt
  arXiv:2108.04828}}].

\bibitem{Dawson:2020oco}
S.~Dawson, S.~Homiller, and S.~D. Lane, {\it {Putting standard model EFT fits
  to work}},  {\em Phys. Rev. D} {\bf 102} (2020), no.~5 055012,
  [\href{http://arxiv.org/abs/2007.01296}{{\tt arXiv:2007.01296}}].

\bibitem{SciencePlots}
J.~D. Garrett, {\it {garrettj403/SciencePlots}}, .
  \url{http://doi.org/10.5281/zenodo.4106649}.

\end{thebibliography}\endgroup
\end{document}